\documentclass[twocolumn,showpacs,amsmath,aps]{revtex4}

\usepackage{graphicx}
\usepackage{dcolumn}
\usepackage{bm}

\begin{document}

\title{Quantum diffraction and interference of spatially correlated photon pairs\\ generated by spontaneous parametric down-conversion}

\author{Ryosuke Shimizu, Keiichi Edamatsu, and Tadashi Itoh}
\affiliation{Division of Materials Physics, Graduate School of Engineering Science, Osaka University, Toyonaka 560-8531, Japan}

\date{\hspace*{3cm}}

\begin{abstract}
We demonstrate one- and two-photon diffraction and interference experiments utilizing parametric down-converted photon pairs (biphotons) and a transmission grating. With two-photon detection, the biphoton exhibits a diffraction-interference pattern equivalent to that of an effective single particle that is associated with half the wavelength of the constituent photons. With one-photon detection, however no diffraction-interference pattern is observed. We show that these phenomena originate from the spatial quantum correlation between the down-converted photons. 
\end{abstract}


\maketitle

The present optical imaging technologies, such as optical lithography, have reached a spatial resolution in the sub-micrometer range, which comes up against the diffraction limit due to the wavelength of light. However, the guiding principle of such technology is still based on the classical diffraction theory established by Fresnel, Kirchhoff and others more than a hundred years ago \cite{Born}. Recently, the use of quantum-correlated photon pairs (\textit{biphotons}) to overcome the classical diffraction limit was proposed and attracted much attention \cite{Boto00,Angelo01}. Obviously, quantum-mechanical treatments are necessary to explain the diffraction-interference of the quantum-correlated multiphoton state. It is well known that much work has been done on two-photon interference using biphotons generated by parametric down-conversion \cite{Rarity90,Ou90,Brendel91,Shi94,Strekalov95}. Although the behavior of two-photon interference can be explained by the standard quantum theory of light, it is also interpreted qualitatively by the concept of the photonic de Broglie wave \cite{Jacobson95}, which is attributable to a special case of the concept of two-photon wave packets \cite{Belinsky94,Rubin94}. Within the concept of the photonic de Broglie wave, the period of two-photon interference is governed by the sum of the momenta, or wave numbers, of the two constituent photons. Recently, the measurement of the photonic de Broglie wavelength of a two-photon state has been experimentally demonstrated \cite{Fonseca99,Edamatsu01}. It has been also proposed that the entangled photons are applicable to novel imaging technologies \cite{Boto00,Bjork01,Abouraddy02}. Boto \textit{et al.} proposed that it is possible to apply entangled photons to high spatial resolution imaging for uses such as optical lithography \cite{Boto00}. According to this proposal, the diffraction-limited spatial resolution of a quantum \textit{n}-photon state is better than that of a single photon by a factor of \textit{n}. D'Angelo \textit{et al.} demonstrated a Young's double slit experiment utilizing parametric down-converted biphotons, and showed that the two-photon diffraction pattern width is narrower, by a factor of 2, than that of classical light \cite{Angelo01}.

In this letter, we present the diffraction-interference patterns of parametric down-converted photons through a transmission grating by both one- and two-photon detection schemes. We also show that the measured diffraction patterns can be explained by the Fourier analysis of the two-photon wave packet, taking account of the quantum correlation between the signal and idler photons. The results also demonstrate a proof of the principle of quantum lithography.

\begin{figure}
\includegraphics[scale=0.27]{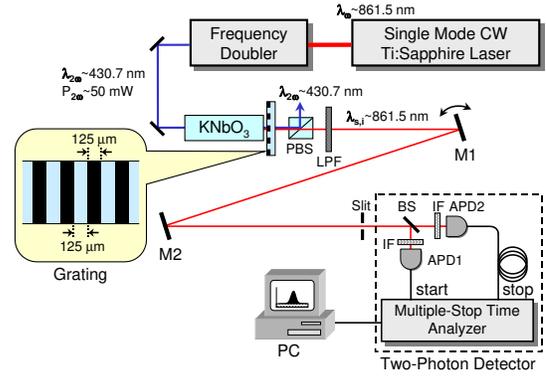}
\caption{\label{fig:exsetup} Schematic experimental setup to observe the biphoton diffraction and interference. PBS: polarizing beam splitter, LPF: long-pass filter, M1-2: mirrors, BS: non-polarizing beam splitter, IF: interference filters, APD1-2: avalanche photodiodes.}
\end{figure}

\begin{figure}
\includegraphics[scale=0.35]{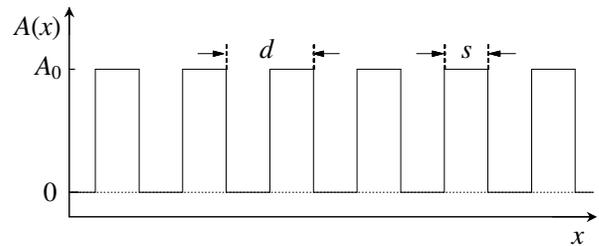}
\caption{\label{fig:Sus} Assumed transmission amplitude through the grating, i.e., the grating function, as a function of the transverse position \textit{x}. $A_0$ is the amplitude transmittance through the grating, and \textit{d} and \textit{s} are the slit period and the slit width, respectively. The number of slits was assumed to be \textit{N}.}
\end{figure}

Figure~\ref{fig:exsetup} shows the schematic drawing of our experimental setup. Pairs of frequency-degenerate photons were generated collinearly by spontaneous parametric down-conversion (SPDC) in a 5-mm-long KNbO$_3$ (KN) crystal pumped by the second harmonic light (50 mW) of a single longitudinal mode Ti:sapphire laser operating at 861.5 nm. The photon pairs were diffracted by a transmission grating (Fig.~\ref{fig:Sus}; slit width: 125 $\mu$m, period: 250 $\mu$m) placed just after the KN crystal. In this geometry, each photon pair passes together through one of the grating slits \cite{Angelo01}. We separated photon pairs from the pump beam by using a polarizing beam splitter (PBS) and a long-pass filter (LPF). By rotating a mirror (M1), we recorded spatial diffraction-interference patterns by a two-photon detector, which consisted of a 50/50\% non-polarizing beam splitter (BS) and two avalanche photodiodes (APD; EG\&G SPCM-AQ161) followed by a multiple-stop time analyzer (EG\&G 9308). To measure two-photon coincidence counting rates, we recorded the number of start-stop events within the time window (1 ns). In addition, we simultaneously recorded the number of start pulses as a one-photon counting rate. In front of each APD we placed an interference filter (IF; center wavelength $\lambda _c$ = 860 nm, bandwidth $\Delta\lambda$ = 10 nm). To compare the results with those of classical lights, we also observed, using the same apparatus, diffraction-interference patterns of the Ti:sapphire laser and that of chaotic light generated from a tungsten-halogen lamp.

It is worth discussing diffraction-interference patterns expected in our experiment. In the previous experiments using Young's double slit, the counting rate for their experiments was obtained by calculating a fourth-order correlation function (superposition method). However, this method seems rather complicated for analyzing arbitrary patterns of two-photon diffraction.
We will now take a semiclassical method using Fourier analysis for a two-photon wave packet, considering the quantum correlation between the constituent photons. This approach corresponds to the Fraunhofer diffraction of the classical optics case. In addition, assuming monochromatic, paraxial, and thin crystal approximation, we need to consider only transverse components of the wave vector \cite{Monken98}. Then, a generalized form for a two-photon wave packet in a one-dimensional system can be expressed by
\begin{eqnarray}
F(q,q')&=&\frac{1}{2\pi}\int dx \int dx' A(x) A(x') \nonumber \\
& & {} \times G(x-x') \exp [i(qx+q'x')],\label{eq1}
\end{eqnarray}
where $q$ and $q'$ are the transverse components of the wave vector, $A(x)$ represents the transmission amplitude through the grating, and $G(x-x')$ represents the transverse correlation between the two photons. The counting rate, which is the same for both two- and one-photon detection, is given by \cite{Burlakov97}
\begin{gather}
R^{(2)}(q,q')=\left|F(q,q')\right|^2,\label{eq2}\\
R^{(1)}(q)=\int dq' R^{(2)}(q,q').\label{eq3}
\end{gather}

First, we assume that $G(x-x')=\delta(x-x')$, that is, a pair of signal and idler photons passes together through the same point \textit{x} of the grating slit. Then Eq.~(\ref{eq1}) can be reduced as follows:
\begin{eqnarray}
F(q,q')&=&\frac{1}{2\pi}\int dx A(x)^2 \exp [i(q+q')x]\nonumber \\
&=&\frac{1}{\sqrt{2\pi}}\mathcal{F}[A^2](q+q'),\label{eq4}
\end{eqnarray}
where $\mathcal{F}[A^2]$ represents the Fourier transform of $A^2$. In this case, the two-photon wave packet is not separable into any product of two independent wave packets. In other words, it is a spatially \textit{entangled} state.
Substituting Eq.~(\ref{eq4}) into Eq.~(\ref{eq2}), we get the biphoton counting rate:
\begin{equation}
R^{(2)}(q,q')\propto\left| \mathcal{F}[A^2](q+q') \right|^2.\label{eq5}
\end{equation}
The two-photon counting rate when we carry out the two-photon detection at the same point $(q = q')$ can be rewritten as
\begin{equation}
R^{(2)}(q,q)\propto\left| \mathcal{F}[A^2](2q) \right|^2.\label{eq6}
\end{equation}
On the other hand, substituting Eq.~(\ref{eq5}) into Eq.~(\ref{eq3}), the one-photon counting rate becomes constant:
\begin{equation}
R^{(1)}(q)=\textrm{const}.\label{eq7}
\end{equation}
This result means that the one-photon diffraction-interference of the biphoton will exhibit no modulation.

Next we consider the case of $G(x-x')=1$, that is, there is no transverse correlation between the two photons. In this case, the Eq.~(\ref{eq1}) can be rewritten as follows:
\begin{eqnarray}
F(q,q') &=&\frac{1}{2\pi}\int dx A(x) \exp(iqx) \nonumber \\
& & {} \times\int dx' A(x') \exp(iq'x') \nonumber \\
&=&\mathcal{F}[A](q)\cdot\mathcal{F}[A](q').\label{eq8}
\end{eqnarray}
In this case, the two-photon wave packet is separable into two independent wave packets. Thus, the two- and one-photon counting rates are given by
\begin{gather}
R^{(2)}(q,q')\propto\left| \mathcal{F}[A](q) \right|^2 \cdot \left| \mathcal{F}[A](q') \right|^2,\label{eq9}\\
R^{(1)}(q)\propto\left| \mathcal{F}[A](q) \right|^2.\label{eq10}
\end{gather}
The one-photon counting rate $R^{(1)}$ corresponds to the classical Fraunhofer diffraction pattern. In addition, when $q=q'$, the two-photon counting rate is just the square of the one-photon counting rate:
\begin{equation}
R^{(2)}(q,q)=\left|R^{(1)}(q)\right|^2.\label{eq11}
\end{equation}
From Eqs.~(\ref{eq10}) and (\ref{eq11}), we understand quite naturally that these results in the case of $G(x-x')=1$ are compatible with classical optics. However, comparing Eq.~(\ref{eq6}) with Eq.~(\ref{eq11}), we see that the results for the biphoton wave packet are quite different from those of two independent photon wave packets in both one- and two-photon counting rates. Biphotons will exhibit half the modulation period that they would have in the classical case in the diffraction-interference pattern for two-photon detection. Moreover, in one-photon detction, no intensity modulation will be exhibited.

\begin{figure}
\includegraphics[scale=0.27]{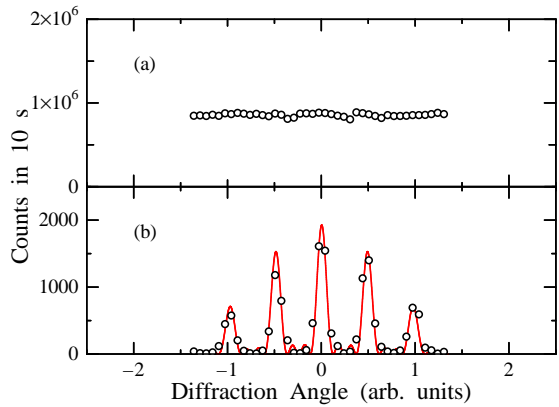}
\caption{\label{fig:PDC} Diffraction-interference patterns of the parametric down-converted photons observed by (a) one-photon detection and (b) two-photon detection.}
\end{figure}

\begin{figure}
\includegraphics[scale=0.27]{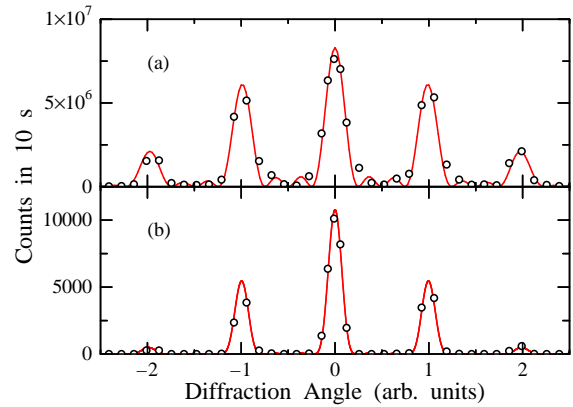}
\caption{\label{fig:TiS} Diffraction-interference patterns of Ti:sapphire laser observed by (a) one-photon detection and (b) two-photon detection.}
\end{figure}

Figures~\ref{fig:PDC} and~\ref{fig:TiS} show the measured diffraction-interference patterns of parametric down-converted photons and the Ti:sapphire laser, respectively. In both figures, the upper graphs represent the interference patterns observed by one-photon detection and the lower graphs are those observed by two-photon detection. The open circles in each figure represent the measured data points. The diffraction angles are normalized by $q=2\pi/d$, where $d$ is the slit period of the grating. These experimental data were fitted with the theoretically expected functions (\ref{eq6})-(\ref{eq7}),(\ref{eq10})-(\ref{eq11}), as indicated by the solid curves. Here, we assume the rectangular transmission profile of the grating as shown in Fig.~\ref{fig:Sus}. Note that in this case the Fourier transform of \textit{A} is
\begin{equation}
\mathcal{F}[A](q)=\frac{A_0}{\sqrt{2\pi}}\cdot\frac{\sin(Ndq/2)}{\sin(dq/2)}\cdot\frac{\sin(sq/2)}{q/2}.\label{eq12}
\end{equation}
One can see that the experimental data are in good agreement with the theoretical prediction \cite{Note}. Especially, the two-photon interference of the SPDC in Fig.~\ref{fig:PDC}(b) exhibits half the modulation period of that of the Ti:sapphire laser in Fig.~\ref{fig:TiS} (see Eqs.~(\ref{eq6}), (\ref{eq10}) and (\ref{eq11})). This result indicates that the biphoton generated by SPDC behaves as an effective single particle that is associated with half the wavelength of the constituent photons. In other words, the diffraction of a biphoton can be explained by the concept of the photonic de Broglie wave. Furthermore, these results manifest the principle of quantum lithography that utilizes the reduced interferometric wavelength of the multiphoton state for optical lithography beyond the classical diffraction limit.
 

It is also noteworthy that the one-photon interference of the SPDC exhibits no modulation, whereas that of the Ti:sapphire laser exhibits normal modulation that can be understood by classical optics. In classical optics, the disappearance of an interference fringe might be caused by the shortage of coherence length owing to the wide spectral distribution of the parametric emission. To be sure that the observed phenomenon originates from a quantum mechanical effect but not from classical effects caused by the spectral width of the light, we also measured a diffraction-interference pattern of thermal light from a halogen lamp. We note that the spectrum of the SPDC emission at the detectors is almost the same as that of the halogen lamp, because we detected both emissions through the same interference filters. However, the measured pattern of the halogen lamp is quite different from that of the SPDC, and is quite similar to that of the Ti:sapphire laser. Thus the disappearance of the one-photon diffraction-interference pattern in the SPDC emission is not due to its spectral width, but to the quantum mechanical effect as described by Eq.~(\ref{eq7}). Together with the two-photon diffraction-interference, the one-photon diffraction-interference of the SPDC also illustrates the non-classical nature of a biphoton that can be explained only by the quantum-mechanical treatment.

\begin{figure}
\includegraphics[scale=0.27]{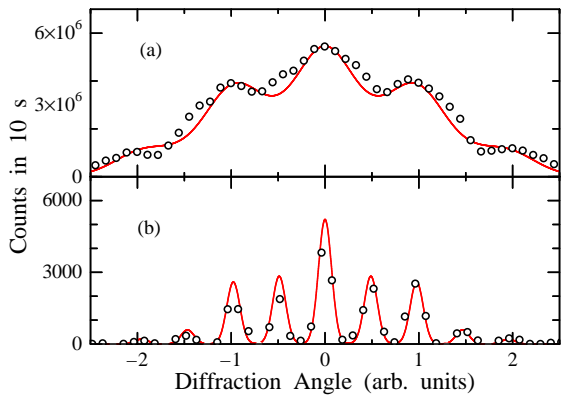}
\caption{\label{fig:CPDC} Diffraction-interference patterns of the parametric down-converted photons observed by (a) one-photon detection and (b) two-photon detection when the pump beam was concentrated.}
\end{figure}

Finally, we consider the case of finite spatial correlation. To achieve this, we observed diffraction-interference by inserting two lenses (focal length = 200 mm) before and after the KN crystal and putting the grating at a distance of 40 cm from the crystal. The increased divergence of the pump beam enlarges the uncertainty of the transverse wave number of the SPDC photons, and this results in the increase of the spatial correlation width between the two photons at the grating \cite{Grayson94}. Therefore, by using the lenses, we can control the spatial correlation between the signal and idler photons. Fig.~\ref{fig:CPDC} shows the observed diffraction-interference patterns. Comparing these results with the former ones shown in Figs.~\ref{fig:PDC} and~\ref{fig:TiS}, we see that the diffraction-interference pattern in the two-photon detection (Fig.~\ref{fig:CPDC}(b)) exhibits the intermediate pattern between the two extreme cases. Furthermore, the biphotons begin to show partial modulation in the one-photon detection (Fig.~\ref{fig:CPDC}(a)), again exhibiting the intermediate pattern between Figs~\ref{fig:PDC} and~\ref{fig:TiS}. Considering a finite correlation width, we can also reproduce these patterns from Eqs.(\ref{eq1})-(\ref{eq3}). Here we assume the Gaussian correlation function:
\begin{equation}
G(x-x')=\exp \left[-\left(\frac{x-x'}{0.56d}\right)^2\right].\label{eq13}
\end{equation}
The calculated patterns are in good agreement with the measured patterns as shown the solid curves in Fig.~\ref{fig:CPDC}. From these results we see that the relation between the extent of the transverse correlation and the object size plays a very important role in forming the diffraction-interference patterns. In terms of quantum lithography, these results indicate that the distribution of a transverse correlation has to be much smaller than an object so as to achieve a spatial resolution high enough to surpass the classical diffraction limit.

In conclusion, we have measured the spatial diffraction-interference patterns of spontaneous parametric down-converted biphotons using transmission grating, and showed that the spatially correlated biphoton exhibits half the modulation period of that of classical light. Also, we have found that one-photon interference of the biphoton exhibits no modulation. Furthermore, by controlling the spatial correlation between the two photons, we have successfully demonstrated that the diffraction-interference pattern changes from the perfectly correlated biphoton case towards the classical case. These experimental results can be understood in a straightforward manner by Fourier analysis of a two-photon wave packet assuming the spatial correlation between the two photons. In this letter, we have discussed the Fourier analysis only for the diffraction-interference patterns formed by the one-dimensional grating, and demonstrated the validity of this analysis. However, this analysis is extendable to arbitrary one- and two-dimensional objects. Therefore, this analysis is very useful for future imaging technology utilizing entangled photons.

This work was supported by the program ``Research and Development on Quantum Communication Technology'' of the Ministry of Public Management, Home Affairs, Posts and Telecommunications of Japan, and by a Grant-in-Aid for Scientific Research from the Ministry of Education, Science, Sports and Culture of Japan.


\begin{references}
\bibitem{Note} From the fitting we found that the ratio of the slit period to the slit width (\textit{d/s}) is 3.3$\sim$3.8, whereas it should be 2 for the actual grating we used. This discrepancy is not due to the quantum mechanical effect, because the same tendency was observed for both biphotons and classical lights. Although the reason for the discrepancy is not clear at present, it might be due to the longitudinal groove pattern of the grating that was printed on a glass plate.
\end{references}
\end{document}